\documentclass[conference]{IEEEtran}

\usepackage{enumitem}
\usepackage{pifont}
\usepackage{tabularx}
\usepackage{makecell}

\newcommand{\xmark}{\ding{55}}%
\newcommand{\cmark}{\ding{51}}%

\IEEEoverridecommandlockouts

\usepackage{cite}
\usepackage{amsmath,amssymb,amsfonts}
\usepackage{algorithmic}
\usepackage{textcomp}
\usepackage{xcolor}
\usepackage{soul}
\usepackage{hyperref}
\usepackage[pdftex]{graphicx}

\begin{document}

\title{Adaptive Webpage Fingerprinting from TLS Traces}

\author{
     \IEEEauthorblockN{
        Vasilios Mavroudis\IEEEauthorrefmark{1}\thanks{\IEEEauthorrefmark{1}Supported by the Defence Science and Technology Laboratory, UK.}
     }
    \IEEEauthorblockA{
        Alan Turing Institute\\
        \textit{vmavroudis@turing.ac.uk}\\
                    }
    \and
    \IEEEauthorblockN{
        Jamie Hayes\IEEEauthorrefmark{2}\thanks{\IEEEauthorrefmark{2}Work done while at University College London, now at Deepmind.}}
    \IEEEauthorblockA{
        University College London\\
        \textit{jamie.hayes.14@ucl.ac.uk}\\
    }
}

\maketitle

\begin{abstract}
In \textit{webpage} fingerprinting, an on-path adversary infers the specific webpage loaded by a victim user by analysing the patterns in the encrypted TLS traffic exchanged between the user’s browser and the website’s servers. This work studies modern webpage fingerprinting adversaries against the TLS protocol; aiming to shed light on their capabilities and inform potential defences. Despite the importance of this research area (the majority of global Internet users rely on standard web browsing with TLS) and the potential real-life impact, most past works have focused on attacks specific to anonymity networks (e.g., Tor). We introduce a TLS-specific model that: 1) scales to an unprecedented number of target webpages, 2) can accurately classify thousands of classes it never encountered during training, and 3) has low operational costs even in scenarios of frequent page updates. Based on these findings, we then discuss TLS-specific countermeasures and evaluate the effectiveness of the existing padding capabilities provided by TLS 1.3.
\end{abstract}

\section{Introduction}\label{sec:intro}

Indeed, past works on \textit{webpage} fingerprinting have demonstrated the feasibility of such attacks~\cite{miller2014know,danezis2009traffic,bissias2005privacy}. This side-channel can be exploited by malicious actors in various settings: by a flatmate residing on the same home network, a coworker using the same Wi-Fi router, another resident in a student dorm, a malicious network administrator, a nefarious ISP, or even a nation state learning about sensitive political affiliations of its citizens. 

The IETF TLS Working Group lists ``privacy'' as a primary goal of TLS 1.2~\cite{rescorla2008transport} and introduced a non-optional record padding  feature in TLS 1.3~\cite{rescorla2018transport}). Padding inserts additional data to obfuscate a record's characteristics. However, the TLS 1.3 specification does not define where, when and how many bytes should be added: ``\textit{Selecting a padding policy ... is beyond the scope of this specification.}''~\cite{rescorla2018transport}.
This is primarily due to our limited understanding of webpage fingerprinting adversaries. Past works have verified the existence of the side-channel but have not challenged sufficiently the scalability of such attacks or studied how distributional shift (phenomenon where the data a model works with changes over time, reducing its accuracy) may affect webpage fingerprinting adversaries and their operational costs. 

\begin{figure}[t!]
	\centering
    \includegraphics[scale=0.89]{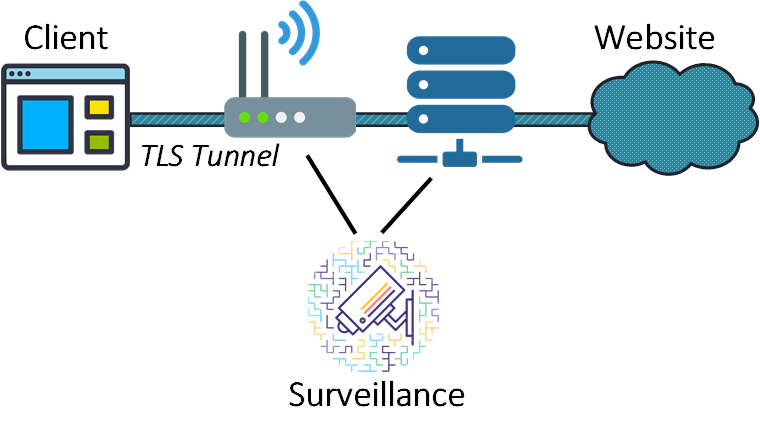}
    \caption{Illustration of a webpage fingerprinting scenario
    where the user loads a webpage through a TLS connection while an
    adversary eavesdrops on the encrypted traffic for surveillance 
    or censorship purposes. TLS conceals the path and the parameters
    of a user request but not the IP addresses of the communicating parties.
    Such a passive adversary can be
    someone sniffing traffic on a local wireless connection, a network 
    administrator, an Internet router operator or the victim’s Internet
    service provider.}
    \label{fig:setups}
\end{figure}

In particular, they have focused on static and relatively small websites (up to $<$ 500 webpages) ~\cite{cai2012touching,danezis2009traffic,bissias2005privacy} without providing reliable proof that these attacks scale further. These beneficial (for the adversary) assumptions (i.e., worst-case scenario for the user) are common practice in the security literature as the accuracy of the attacker serves as a privacy upper-bound that the user can reliably assume under all possible circumstances. However, such ``optimal'' scenarios do not provide strong evidence of a model's efficiency in realistic, non-ideal conditions. 
For example, a fingerprinting model that is accurate under some very specific optimal conditions (e.g., target set of 100 static webpages) is not guaranteed to remain as performant in all settings. If the set of potential webpages is larger or the contents of the pages change frequently these attacks are likely to perform worse. This leaves a gap in our understanding of those attacks and potentially creates ambiguity about the degree of threat such attacks pose to the TLS protocol. 

In this work, we shed light on the capabilities of modern webpage fingerprinting adversaries and use our insights to inform defence strategies that make optimal use of TLS record-padding (and other possible countermeasures). We first compile a list of important factors (e.g., scale, distribution shift) that can directly affect the performance of such an attack and, consequently, its practicality. We then outline the basic properties that a TLS fingerprinting adversary should have in order to be practical and investigate if existing webpage fingerprinting techniques meet them. We find that these previous techniques were designed to operate on a target set under static, non-changing conditions (e.g., constant webpage contents) and lack the ability to rapidly and inexpensively adapt to changes in the target webpages. To answer if a realistic and performant adversary under non-optimal conditions is possible, we introduce \textit{adaptive fingerprinting}. 

Our design is based on learning low-dimension representations of input data, referred to as \textit{embedding models}, which have been the basis for recent advancements both in natural language processing (NLP) and computer vision (CV) (e.g., FaceNet~\cite{schroff2015facenet}, BERT~\cite{devlin2018bert}). Their ability to process and preserve properties of high-dimension data allows practitioners to train downstream tasks with relative ease (e.g. question answering in NLP, sentiment classification in NLP, multi-label image classification for CV). Following from~\cite{hayes2016k,sirinam2019triplet} that focused on Tor, we show that such advancements can be readily applied in the TLS traffic analysis space. By utilizing embedding models we can retain high detection accuracy in larger data settings and allow for rapid and inexpensive adaptation to distributional shift without the need for retraining.

Based on these findings, we confirm that neither TLS 1.2 nor TLS 1.3 can reliably provide privacy in the presence of webpage fingerprinting adversaries, even in the case of websites with thousands of pages. Consequently, users can rely on the TLS protocol to protect private information (e.g., credit card numbers, medical results) but not their browsing habits (e.g., shopping websites, online encyclopedias, medical websites, social services). 

\section{Preliminaries}

\subsection{The Transport Layer Security Protocol}
The TLS protocol is a cryptographic protocol that is commonly used to establish secure two-party communication channels over untrusted networks (e.g., the Internet). It is utilized in a wide range of applications such as web browsing, email, instant messaging and voice over IP, and employs end-to-end encryption between the two parties to protect the integrity and the confidentiality of the transmitted data. The two participants first negotiate the ciphersuite details and then perform an one-time \textit{handshake} to generate the cryptographic keys that will be used to protect the contents of their communication. Following a successful handshake, all the data exchanged is encrypted. Note, however, that the IP addresses of the communicating parties are not concealed. Moreover, to prevent man-in-the-middle attacks a client accessing a TLS-enabled server verifies the identity of the server through a public-key certificate issued by a trusted certification authority. For a detailed analysis of the TLS protocol please refer to~\cite{felt2017measuring,arfaoui2019privacy,rescorla2018transport, dowling2016cryptographic}. In this work, we focus on the latest two versions of TLS, 1.2~\cite{rescorla2008transport} and 1.3~\cite{rescorla2018transport}.

\subsection{Webpage vs. Website Fingerprinting}
As discussed in Section~\ref{sec:intro}, the majority of past works has focused on 
website fingerprinting against users that route their traffic through anonymity
networks (e.g., the Tor anonymity network). Such works aim to uncover
the \textit{website} visited by the user from a pool of possible websites that
are of interest to the eavesdropping adversary. Webpage fingerprinting
is orthogonal to that goal as it aims to identify the specific webpage 
accessed by the user. So far, webpage fingerprinting has not
received as much attention in the literature, despite the fact that 
the Tor user-base is only a fraction of the total number of TLS users. 

From a technical perspective, both attacks rely on extracting
data-transmission patterns (i.e., byte counts, sender and recipient) to uniquely 
identify a website or a webpage. However, \textit{webpage} fingerprinting 
presents additional challenges, as websites tend to reuse the same
template/theme in all their pages. Thus, webpages belonging to the same website
exhibit only partially unique transmission patterns, with the only
differentiating factor being the content of each page. This limits 
the amount of useful identifying information one can extract
from the traffic stream. In contrast, in website fingerprinting, 
the whole stream can be uniquely-identifying as websites usually 
use different themes/template.

We believe that webpage fingerprinting is a pressing issue that has
received disproportionately low attention. Especially, when considering
the number of users that are exposed to such attacks and the nature 
of the data that can be leaked (e.g., health info from users browsing 
condition-specific articles on medical websites). 
In comparison, website fingerprinting attacks affect mainly Tor users 
(perhaps a more sensitive group admittedly) and reveal only the website
visited, which in cases of large websites (e.g., Wikipedia) may not leak 
much information about the user's interests or habits.

\section{Adversarial Setup}\label{sec:adversaries}
In this section, we introduce the threat model, the attack scenarios and the practicality constraints that we will consider in the rest of this work.

\subsection{Threat Model}\label{sec:threatmodel}

We assume a polynomially-bound \textit{passive} adversary that can capture (but not tamper with) the packets exchanged between the client and the server. The client communicates with a server over an encrypted channel established through TLS while the adversary intercepts some or all of the packets exchanged. 
We assume that the adversary can observe network traffic without any additional encryption applied other than what is done so by TLS, and without any more information about the traffic other than what TLS reveals. 
While in certain cases there may be some additional encryption layers (e.g., Wi-Fi encryption), our adversary is assumed to be able to circumvent it (e.g., WPA-PSK 4-way handshake de-authentication attack for users sharing the same network, resides on the ISP level). 
We argue this is a reasonable assumption to make for committed adversary, as this is precisely the reason TLS uses end-to-end encryption  -- network traffic transmitted over the Internet is exposed to eavesdroppers.

Such on-path adversaries are standard in traffic fingerprinting literature~\cite{danezis2009traffic,bissias2005privacy,dubin2017know,wang2014effective,panchenko2016website,sirinam2018deep, juarez2014critical, sirinam2019triplet,cai2012touching,hayes2016k,bhat2019var,wang2016realistically, DBLP:conf/uss/ShustermanKHMMO19}. 
This broad threat model allows us to consider a wide range of potential adversaries: the adversary could be a flatmate residing on the same home network, a coworker using the same Wi-Fi router, another resident in a student dorm, the network administrator in a company, or even a nefarious ISP acting independently or co-operating with an adversarial nation state (Figure~\ref{fig:setups}). 
Webpage (and website) fingerprinting attacks are passive and thus hard to detect at any point on the communication path (which is important in many surveillance scenarios).

The goal of the adversary is to infer the specific \textit{webpage} visited by the user (e.g., Wikipedia article, eBay product page). As neither TLS 1.2 nor 1.3 conceal the IP address of the webserver, we assume that the adversary is aware of the \textit{website} that 
the user is visiting. In particular, the adversary can observe the IP addresses of all the servers involved in a page load. Although an IP address may correspond to many websites (i.e., multihosting, CDN hosting), this is neither guaranteed (e.g., large websites have dedicated servers) nor should be relied upon to provide a provably large/secure anonymity set. In their current form, the privacy protections that may be provided by such platforms will be a by-product rather than a reliable security property that is guaranteed for all the webpages they serve. Note also that our threat model does not cover VPN services where the IP addresses of the servers/websites visited are concealed.
The above threat model is in line with the adversarial setup outlined in the specifications of the TLS protocol versions 1.2 and 1.3~\cite{rescorla2008transport,rescorla2018transport}.

\subsection{Realistic Fingerprinting Scenarios}\label{sec:scenarios}
We now focus on fingerprinting scenarios that 
provide a realistic representation of the conditions under
which an adversary has to operate. While this may make it
harder to design and implement effective attacks, it enables us to
draw reliable conclusions about the capabilities of the adversary in 
practical settings. In particular, we focus on three aspects of such scenarios: 1) Number of
classes (e.g., webpages, websites), 2) Distributional shift (e.g., content updates),
and 3) Shared resources (e.g., common HTML theme, shared images).

\subsubsection{Number of classes}
Past works on webpage fingerprinting considered scenarios where the user is assumed to visit a \textit{fixed} set of known webpages while the adversary aims 
to infer which webpage was loaded~\cite{cai2012touching,danezis2009traffic,bissias2005privacy,miller2014know}.
Unfortunately, their experiments were conducted on datasets of up to 500 webpages. Such datasets have been criticized as being unrealistically small~\cite{sirinam2018deep} 
and led to doubts about the practicality of the proposed attacks, especially as many modern websites include hundred or even thousands of unique webpages.
In comparison, recent works on website fingerprinting evaluated their proposed techniques to significantly larger sets (a few thousand websites) and showed that adversaries achieve a high performance under ideal conditions~\cite{sirinam2018deep,bhat2019var}. Overall, we argue that fingerprinting techniques should be evaluated in at least one scenario with a moderate or large number of classes. 

\subsubsection{Distributional Shift}
Another common assumption in past works on fingerprinting is \textit{static} webpage contents. While assuming content invariability may look reasonable at first glance, it results in significant performance degradation in practice as pages change~\cite{juarez2014critical}. A model that is trained to classify a set of pages (e.g., Wikipedia articles, subreddits, eBay listings) will have to retain its accuracy as their contents get updated. This can be achieved either by retraining the classification model on the latest version of the webpages or through other means. From an adversarial perspective, the cost of keeping up with the ever-changing contents is directly connected to the practicality of the technique. For example, a model that needs to be retrained each time one or more webpages get updated is likely to incur large operational costs thus making the technique impractical, even if it achieves high accuracy. Overall, the degree of tolerance to distributional shift and the cost of adapting to changes are also important factors that must be considered when evaluating a fingerprinting technique.
The rate of distribution shift will differ from website to website, for example the landing page of Reddit changes at much higher frequency than the landing page of Wikipedia. We will see in Section~\ref{exp:acrossclasses} that our proposed attack can operate under extremely challenging settings with high distributional shift, where the set of monitored webpages that have been used to train the attack share no overlap with webpages that are monitored during the evaluation phase.

\subsubsection{Shared Resources}

It is common for the pages of a website to share a HTML theme (e.g., the same stylesheet, Javascript imports, background image files). 
This reduces the volume of unique information transferred in each page load, thus making it harder for the adversary to uniquely identify each webpage. 
Scenarios should also account for cases where only part of the content is unique.

\subsection{Practicality Considerations}\label{sec:practicalconsiderations}
We now introduce a list of requirements for a fingerprinting technique to be considered practical and realistic.

\subsubsection{Accuracy \& Scalability}
An effective fingerprinting technique needs to provide high inference accuracy
for at least medium-sized and preferably large-sized websites (with regards to their number of webpages).
For example, a technique that achieves 80\% accuracy on a set of 100 webpages is not necessarily 
equally accurate when used on larger sets.

\begin{figure*}[t]
	\centering
    \includegraphics[scale=0.80]{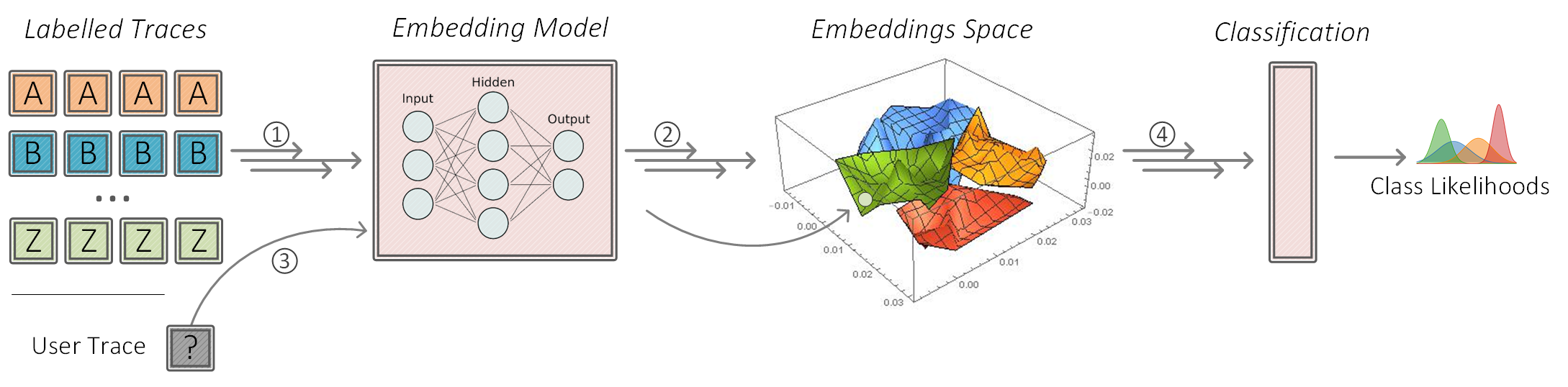}
    \caption{The eavesdropping adversary maintains a dataset of \textit{labeled traces} from the webpages they monitor. These traces are processed by the \textit{embedding neural network} and form the set of reference points. The reference points are then
    used to classify the user's traffic based on a proximity-based algorithm (e.g., k-nearest neighbours). Optionally, the adversary can keep populating the dataset with new reference points to stay up-to-date with the latest version of the webpages, without the need to retrain the embedding model.}
    \label{fig:pipeline}
\end{figure*}

\subsubsection{Adaptability}
As discussed in Section~\ref{sec:scenarios}, websites periodically add new webpages or update the contents of existing ones. Practical fingerprinting techniques must be resilient to such distributional shift and retain their accuracy~\cite{juarez2014critical}. 
Moreover, while adversaries may be able to cope with small page updates, it is not uncommon
for webpages to have most of their content gradually replaced (through small but frequent updates).
This gradual process leads to a large distributional shift where the current version of a page
has a very small overlap with the version the model was initially trained on. The practicality and the performance of a fingerprinting technique
depend on its ability to adapt to such changes (e.g., frequent retraining, low generalization error) and the operational cost this entails.

\subsubsection{Provisioning \& Operational Costs}
Making inferences from traffic traces should come at a reasonable operational cost
(i.e., in time and computational resources), while provisioning the fingerprinting model
may have a larger one-off cost. Minimizing these costs results in more practical 
and easily-applicable models

\subsubsection{Version-agnostic}
While past works have focused on a specific protocol version, it is 
advantageous for a practical adversary to be able to fingerprint webpages regardless
of the underlying protocol version used by the user. For example, a fingerprinting
deployment that is tailored to only one protocol version of the TLS protocol could 
potentially be temporarily circumvented by switching to a different version (e.g., from TLS 1.2 to 1.3)
or even to a different ciphersuite entirely. This is not a strict requirement (protocol-specific attacks 
can be also very effective) but we consider this a desirable (albeit not necessary) feature for highly-transferable models.

\section{Adaptive Fingerprinting}\label{sec:adaptive}

Our proposed methodology allows adversaries to fingerprint webpages from non-static, changing webpages. The core components of our system (Figure~\ref{fig:pipeline}) are the embedding neural network and the classification algorithm that attributes samples to classes (i.e., traffic traces to webpages). Its operation comprises of three processes: \textit{provisioning}, \textit{fingerprinting}, and \textit{adaptation}. 

The computationally demanding provisioning process takes place only once, while the lightweight fingerprinting and the adaptation processes are executed iteratively throughout the lifecycle of the deployment. This is primarily possible due to the generic nature of the \textit{embeddings} generated as part of the \textit{mapping} step (Section~\ref{subsec:fingerprinting}). The following sections provide the details of these operations.

\begin{figure*}[t]
    \center
    \includegraphics[scale=1.4]{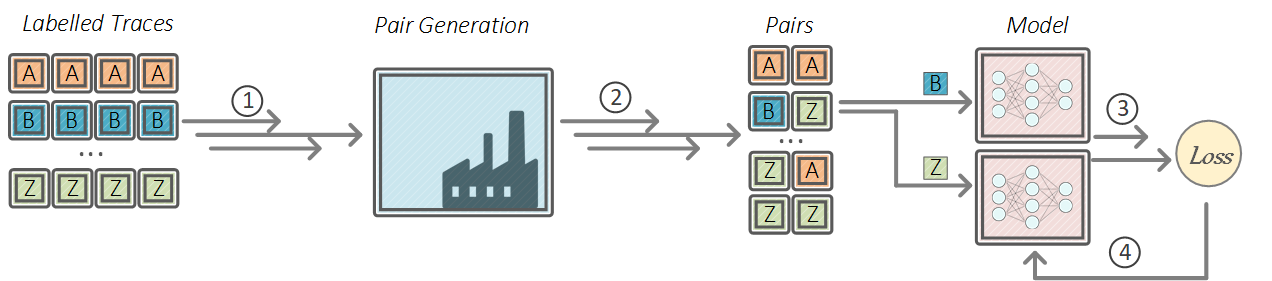}
    \caption{To train the embedding model, we use a dataset of labeled traffic traces that originate from the same website (e.g., Wikipedia). Using that set, we generate pairs of traces from the same class and from different ones (i.e., positive and negative pairs). These pairs are then used to iteratively train the model until sufficient accuracy has been achieved.}
    \label{fig:training}
\end{figure*}

\subsection{Provisioning}\label{sec:provisioning}
Before the system is usable, the embedding neural network that reduces the dimensionality of the
input traffic traces needs to be trained. Our training process is illustrated in 
Figure~\ref{fig:training} and involves four steps.

\subsubsection{Data Collection \& Preprocessing.}
\begin{figure}
    \center
    \includegraphics[scale=1.1]{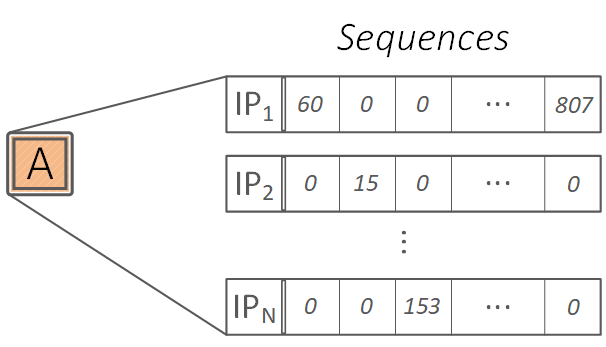}
    \caption{Illustration of how a network traffic of a pageload (labelled ``A'') is converted into IP sequences. Websites often load various parts of their pages (e.g., JavaScript files, images) from  different servers (e.g., for load balancing). Thus, each time the webpage is loaded, the client establishes TLS sessions with and fetches content from \textit{several} different servers. Each sequence corresponds to the bytes sent by one of these servers while the first sequence always corresponds to the user.}
    \label{fig:sequences}
\end{figure}

Initially, the adversary compiles a list of webpages
and then proceeds to repeatedly load each webpage several times. 
For each visit, the network traffic between the client and the server is stored in a packet
capture file (pcap file) and placed in a library of \textit{raw} traces.
Following the collection of the raw traffic traces, the adversary processes them into
sequences of integers (Figure~\ref{fig:sequences}). Each sequence corresponds to one 
of the IP addresses that transmitted data during the pageload and contains the 
byte-counts sent by that IP address over time.

In particular, each time an IP address sends out traffic, the new byte-count is appended to the corresponding sequence while the rest of the sequences are appended with a zero-count 
element. This is done to preserve the relative order of the transmissions. When an IP address sends more than one consecutive packets (i.e., no traffic from other IP addresses is interleaved),
the byte-counts of those packets are aggregated and only their sum is appended to the sequence. 

Unlike our approach, prior works represent the data exchange as a single sequence where incoming packets are denoted by their byte-count and a negative sign, while outgoing 
by the byte-count with a positive sign~\cite{panchenko2016website,hayes2016k,bhat2019var}. This is equivalent to using only two IP sequences, one for incoming and one for outgoing traffic.
The reduction in the number of sequences is because anonymity networks (e.g., Tor) conceal the IP addresses involved in a pageload as all the traffic is routed through an entry node of the network.
In contrast, TLS does not protect the IP addresses of the servers involved in a page load (e.g., user's client, main Wikipedia server, servers for auxiliary JavaScript files and images).
Following this step, the sequences can be optionally \textit{quantized} to eliminate noisy artifacts (e.g., small differences in the byte counts). At the end of this
process, the adversary has a dataset of labeled traces (each trace is a set of IP sequences corresponding to a single page load) that can be used to train the neural 
network (leftmost block in Figure~\ref{fig:training}).

\subsubsection{Pair Generation}
Given the dataset of labeled traces, the adversary generates \textit{positive} and \textit{negative pairs}. Positive pairs comprise of two traces corresponding to the 
same webpage, while negative pairs to different ones. The most straightforward strategy to generate pairs is at random, while more advanced techniques have been also proposed in the relevant ML literature (e.g., Hard-Negatives, Semi-Hard-Negatives~\cite{wu2017sampling,harwood2017smart,schroff2015facenet}). The pairs are labeled based on the similarity of the samples (1 for similar, 0 for different) and are then used to train the embedding model.

\subsubsection{Training}
In this step, we train the machine learning model to produce embeddings that are in close proximity when the input traces originate from the same webpage, and far-apart otherwise. 
Intuitively, the role of the \textit{embedding network} is to extract robust features that are less sensitive to artifacts (e.g., packet re-transmissions, non-deterministic resource loading order) and map the samples in the \textit{embedding space} (Figure~\ref{fig:pipeline}). Classification algorithms (e.g., k-nearest neighbours) that rely on the distance between the samples (e.g., Euclidean, cosine) perform significantly better in low-dimensional spaces compared to when they operate on the original high-dimensional feature space~\cite{snyder2017deep,golovko2007dimensionality,906006,van2009dimensionality}. The specific architecture of the neural network and its training details depend on the needs of the adversary and the use case.

Following the methodology outlined in~\cite{hadsell2006dimensionality,chopra2005learning}, for every training pair, we embed the two input sequences and compute the \textit{similarity} of the two embeddings. For positive pairs, the similarity must be approximately equal to $1$, while for negative pairs approximately equal to $0$. To estimate the correctness of our model and update the network parameters accordingly, we compute the \textit{contrastive loss}~\cite{chopra2005learning} given by the formula:

\begin{equation}\label{for:contrastiveloss}
\mathcal{L}(d, y) = yd^2 + (1-y) \max(margin-d,0)^2
\end{equation}

where $d$ is the (Euclidean) distance between the two embeddings $e_1$ and $e_2$ ($d = ||e_1 - e_2||_2$), $y$ is the known similarity label of the pair and the $margin$ is a user defined parameter (a scalar) used to improve the separation between the different classes in the embedding space (i.e., dissimilar pairs should have a distance at least equal to the \textit{margin}). A larger \textit{margin} improves the robustness of the features extracted by the model (i.e., ensures a separation between the embeddings of samples from different classes) but large values can prevent the model from learning at all. The training process is completed once sufficient performance has been achieved and produces a model that can determine if two traffic sequences originate from the same website based only on the leakages of the cryptographic protocol used. {Given a sufficiently large dataset, the sheer number of possible positive and negative pairs prevents the model from overfitting (e.g., memorizing class-specific patterns) and promotes learning to extract information-rich low-dimensional representations.}

\subsubsection{Initialization}
Following the training of the embedding model, the system is populated with data that 
serve as reference points when classifying unlabeled traffic traces captured by the adversary.
The adversary compiles a list of the webpages they intend to fingerprint, crawls them and 
and embeds the traffic sequences to generate a \textit{reference set} of labeled embeddings
(steps 1 and 2 in Figure~\ref{fig:pipeline}). The reference set is then stored and 
used every time an unlabeled traffic trace is classified.

\subsection{Fingerprinting}\label{subsec:fingerprinting}
Given an initialized deployment with a populated \textit{reference set},
the adversary can then proceed to fingerprint unlabeled samples captured from the user's traffic.

\subsubsection{Capturing and Mapping}
Depending on the setup, the adversary may capture the user's traffic 
at an Internet service provider (ISP) level or may reside in the same
network and thus capture the traffic locally.
Upon converting the packet capture into sequences,
the adversary uses the embedding model to map the unlabeled sequence into
the embedding space (step 3 in Figure~\ref{fig:pipeline}).
The embeddings generated for each sequence are continuous vectors that represent the packet exchange in a low-dimensional space. It should be noted that, while this step
determines the spatial proximity of the embeddings (based on their characteristics),
the process is completely label-agnostic. This provides greater flexibility to the
whole system as the embedding model does not need to be retrained if the labels change. 
In contrast, the majority of past works perform both the feature-extraction and the 
classification through the same model (e.g., convolutional neural networks~\cite{sirinam2018deep}),
thus fitting is specific to the labels seen during the training. This is an important difference
with past works as it minimizes the memorization of the specific characteristics of the 
webpages in the training set. In Section~\ref{sec:experiments}, 
we examine how accurately the embedding model can map sequences from webpages never seen during training.

\subsubsection{Classifying}
The adversary then classifies the embedding that corresponds to the user's traffic trace (step 4 in Figure~\ref{fig:pipeline}). Intuitively, each captured sample is classified based on the labeled traces (reference points) 
that are in its proximity in the embedding space. The distance metric and the 
classification algorithm can be freely chosen by the adversary.
In most cases, the algorithm outputs a list of the most probable labels for the examined sample
and the frequency each one of them occurred (i.e., number of samples in proximity with 
that label).

\subsection{Adaptation}\label{subsec:adaptation}
Besides the initialization and the fingerprinting processes, our methodology involves an 
optional adaptation process. It provides a computationally lightweight process that brings the deployment 
up to date with changing webpages and prevents performance degradation~\cite{juarez2014critical}.

Initially, the adversary crawls and identifies the webpages that have been 
updated. The adversary can sequentially visit the webpages or in cases of larger websites, 
use techniques for monitoring and detecting changes in millions of webpages that were
originally developed for web-archiving purposes~\cite{tan2007efficiently,spaniol2009catch}.

Given one such page, the adversary loads it, collects a traffic trace and fingerprints it
as outlined in the previous section. If the accuracy of the classifier
is not adequate, the adversary crawls the page several times
and updates the labeled traces in the reference samples dataset.
The decision to update the reference samples of a particular class (in case the 
contents of the page have changed) can be taken based on a user-defined accuracy
threshold (e.g., maximum discrepancy from the accuracy of the freshly-initialized deployment). 

The main advantage of this process is that it does not require any retraining of the
model nor of any other component of the system (unlike the majority of past works on fingerprinting~\cite{wang2014effective,panchenko2016website,hayes2016k,bhat2019var,sirinam2018deep,danezis2009traffic,miller2014know,cai2012touching}). This is because training in pairs prevented the embedding model from overfitting on the training set and pushed it to generate effective low-dimensional embeddings even for traces from webpages not encountered during training.

Retraining a machine learning model is a costly operation and would impede the scalability 
of the attack if it was to be executed every time one of the thousands of pages/websites is updated.
Instead, adaptive fingerprinting enables the adversary to remain up to date with 
fast-changing webpages through a short sequence of inexpensive and low-complexity operations.
\section{Datasets}\label{sec:datasets}
To better understand the performance of fingerprinting adversaries under non-optimal conditions, we evaluate our proposed fingerprinting technique by introducing two new datasets with TLS traffic traces: \textit{Wiki19000} with 19,000 classes for TLS 1.2, and \textit{Github500} with 500 classes for TLS 1.3. 
To the best of our knowledge, there are no other publicly available datasets of this size with TLS 1.2 and 1.3 traces.

As outlined in Section~\ref{sec:intro}, our goal is to enable further research into (adaptive) adversaries, scalability and webpage fingerprinting. For this purpose, we publicly release both our datasets as well as our trained models. However, in order to limit potential abuse of our published data and models, we sought to crawl websites that:
\begin{itemize}[leftmargin=*]
    \item Do not have sensitive content (e.g., medical websites).
    \item Explicitly allow crawling (e.g., ``crawl-delay'' directive in robots.txt).
    \item Use a common HTML theme in many pages with varying content.
\end{itemize}

We identified \textit{Wikipedia} and \textit{Github} as services that fulfill the above requirements: Both websites have a large number of webpages that use the same theme but the text and media contents varying significantly between the webpages. Both websites permit crawling and their contents are generally not privacy-sensitive. We manually removed entries on topics that are more likely to be monitored~\cite{RimmerPJGJ18,rahman2020utility}) to make it harder for adversaries to abuse our datasets, trained models and source code. In contrast, targets such as Amazon, eBay and Reddit do not permit crawling and public fingerprinting models trained on these websites have a high abuse potential.

\subsection{Technical Details}
Each dataset contains (encrypted) traffic traces as they would be captured by the eavesdropping adversary introduced in Section~\ref{sec:threatmodel}. We employed 100 Amazon EC2 instances distributed over five geographical regions (20 instances in each region).  We opted for a small instance type, which features 2 GBs of RAM and up to 5 Gbps network bandwidth.

Each instance crawled the same list of URLs, captured the generated traffic by each of the independent pageloads, stored it as a pcap file and processed it into sequences of bytes (Figure~\ref{fig:sequences}). To automate the crawling process, we used Python 3.7 with the Selenium automation framework (https://selenium-python.readthedocs.io/). To determine the browser to be used with selenium, we ran a small-scale experiment that did not indicate significant differences in the captured traces between Chrome and Firefox. However, instances using Firefox exhibited decreased stability. For this reason and due to the substantial difference in their market shares, we opted to use Google Chrome.

Each instance ran only one crawling process that visited each URL on the list sequentially in a random order. Before each visit, the crawler launched a Tcpdump~\cite{fuentes2005ethereal} process and then proceeded to load the page with Google Chrome. Upon waiting 10 seconds for the contents to fully load, the Tcpdump process was terminated and the captured traces were stored on a pcap file.

\noindent\textit{Caching.}
The instances loaded the webpages strictly sequentially, and in addition to this, we made sure that there were no prefetched resources, history or caches. This is to simulate a user that uses the ``incognito'' option of their browser. No page loads took place in non-incognito browsing mode to prevent artifacts in our traces from cached favicons~\cite{solomospersistent}. 
We argue that evaluating the attack against an Internet user who exclusively uses ``incognito'' mode for sensitive browsing is not a particularly restrictive assumption. Prior work has shown that users visiting sensitive webpages often use ``incognito'' mode so that browsing does not leave traces on their computer (e.g., history, cookies)~\cite{DBLP:conf/chi/Abu-SalmaL20}. Moreover, caching has been found to assist such attacks in a previous version of chrome (c.f. \cite{pironti2012identifying}).

\noindent\textit{Multiple requests.}
Our datasets include a single independent trace for each webpage load. In a real browsing session, however, a user might load several webpages sequentially. If these webpages are fetched from different websites, the adversary can easily discern each load based on the IPs involved. 
Alternatively, if the webpages loaded are from the same website (e.g., when the user is reading the first page before loading the second), an adversary can actually enhance the accuracy of their inferences: Past works on webpage fingerprinting~\cite{danezis2009traffic,miller2014know} have identified that the pages loaded during a browsing session are not independent. This is because the link structure of a website guides the browsing sequence. Based on this realization, they experimentally proved that a webpage fingerprinting model (trained to identify individual webpages) can pass its inferences to a hidden Markov model (trained on the website's link graph) and substantially boost the adversary's accuracy.

\subsection{The Wikipedia dataset}
Our Wikipedia dataset (\textit{Wiki19000}) consists of encrypted traffic traces from 19,000 distinct Wikipedia articles. We randomly chose 20,000 Wikipedia webpages and removed stub articles and indexing pages, as well as a few articles on potentially sensitive topics (e.g., 1989 Tiananmen Square protests). The remaining ${\sim}$19,000 webpages were placed in a list to be used by the crawlers. To diversify our traces, each crawler shuffled the list and visited each article only once in a random order. The crawling process lasted approximately three days and cost approximately \$300, thus making it relatively inexpensive to replicate our data collection and further extend the dataset. 

Wikipedia uses TLS 1.2 and the page contents are usually loaded from two servers (one 
for text content and another one for media resources). We examined the contents of the 
Wikipedia articles crawled over the period these three days and found only minor changes on some articles. In total, the resulting dataset contains 1,900,000 traffic traces (100 traces for each URL). Capturing 100 samples per class is on the lower end and is consistent with recent works~\cite{sirinam2019triplet}. The size of the dataset is substantially larger that any other website fingerprinting dataset available in public. An adversary who can reliably  fingerprint this dataset can potentially also train their model against other websites with equal or fewer webpages (e.g., local city portals~\cite{URBAN200249}, user directories~\cite{DBLP:conf/bigdataconf/JayawardanaNJ0J20,jati2011web}, government websites~\cite{walia2010government}, health information resources~\cite{medline2022}).

Moreover, we also include 50,000 samples from Wikipedia with TLS 1.3 (500 Wikipedia webpages with 100 samples each).

\subsection{The Github dataset}
For our second dataset (\textit{Github500}), we chose Github as it was one of the few websites that had deployed TLS 1.3 at the time of the data collection and permits crawling of its pages. Moreover, it features a moderate number of webpages (i.e., projects) all sharing a common HTML theme.

Github allows projects to display a README page with information on the project as well as with installation and usage instructions. The overlaying Github template is common for all the projects but the contents of each page are managed by the project's contributors. Such pages include text, images and sometimes videos. Images and videos are stored either internally on Github or on external servers. Our dataset was generated by visiting the top 500 Github project pages (https://gitstar-ranking.com/repositories), 1,000 times each. Each crawler instance shuffled the list of URLs and then visited each Github page 10 times over the span of several hours. We chose to use the top-500  projects, as actively maintained projects with substantial contributions almost always have a detailed README page with information. In contrast, a random selection of pages (similar to that of Wiki19000) gave us mostly README pages with either no or minimal content (e.g., a single command line to compile the project).

Github uses TLS 1.3 and exhibits increased variability across various dimensions.
It employs a significantly distributed infrastructure and
advanced load balancing techniques causing various discrepancies between subsequent 
pageloads of the same page. Moreover, the number of servers involved 
is heavily dependent on the contents of each project page (e.g., externally hosted
images, scripts and media). Due to this variability of the traffic patterns,
we opted to collect 1,000 traces per class (in line with~\cite{sirinam2018deep}).
The dataset contains 500,000 traffic traces: 500 articles visited in random order 
10 times by 100 crawler instances. Similarly with Wikipedia, we observed that Github
project pages were not updated frequently (e.g., on an hourly basis) nor radically as 
they mostly provide compilation and usage details.

\section{Experimental Evaluation}\label{sec:experiments}
In this Section, we evaluate our proposed methodology by deploying and testing its performance 
on real data. We use three scenarios that simulate real-world fingerprinting setups with non-optimal 
conditions for the adversary. We focus primarily on webpage fingerprinting scenarios as such attacks 1) have been systematically overlooked in the literature (cf. website fingerprinting attacks),  2) are more severe as 
they can affect many more users (i.e., the number of Tor users compared to that of Web users) and 3) pose a 
more pressing threat to the privacy of individuals. For example, a website fingerprinting attack could infer 
that the user is visiting Wikipedia, while webpage fingerprinting attacks uncover the exact article loaded.

As discussed in Section~\ref{sec:intro}, past works have already shown that modern machine learning techniques can achieve very high accuracy~\cite{wang2014effective,hayes2016k,sirinam2018deep}. Thus, our primary focus is to study whether an adversary can retain such a high performance in a considerably larger scale setting while simultaneously alleviating the need for static targets.

\textit{Top-N Accuracy.} 
We measure success of the attack with respect to a top-$n$ success metric. 
A top-$n$ adversary is considered to win the fingerprinting game if the true label is contained within the top $n$ predicted labels. Top-$n$ accuracy is a common metric used in computer vision -- we argue that it is also a useful measure of leakage in adversarial traffic analysis settings by way of the following example: 
an adversary captures a user’s TLS trace loading webpage A (true label) from a website with 1,000 webpages. The adversary predicts that the trace is from either webpage A or B with a probability 49.9\% and 50.1\% respectively (and assigns 0\% to all other webpages). Even though A was not the top-predicted label, clearly a lot of information about the true webpage (A) has been leaked to the adversary.

\subsection{Implementation \& Parameterization}
For the implementation of our neural network, we use the Python 
deep learning library Keras~\cite{gulli2017deep} as the front-end, and
Tensorflow~\cite{199317} as the back-end. For the data preprocessing
and classification algorithm, we use Numpy~\cite{oliphant2006guide} and Scipy~\cite{virtanen2020scipy}, respectively.

As outlined in Section~\ref{sec:adaptive}, we use a \textit{contrastive loss}~\cite{chopra2005learning} to train our model on both positive and negative pairs. 
The \textit{margin} of the loss function was set to be $10$ and was determined through grid search (\cite{bergstra2012random,bergstra2011algorithms}) among smaller and larger values. To measure the proximity of the traffic embeddings, we use the Euclidean distance. The sizes of the hidden layers and the dimensionality of the produced embeddings were determined through grid search (see Table~\ref{tbl:model}). The architecture of the embedding model and its hyperparameters were chosen so as to maximize the fingerprinting performance and accuracy.

\begin{table}
\centering
\resizebox{0.95\columnwidth}{!}{
\begin{tabular}{l  l }
\textbf{Hyperparameter} & \textbf{Value(s)}\\
\hline

\emph{Input layer} & 30 LSTM units\\ 
\emph{\# hidden fully connected layers} & 4 layers \\ 
\emph{Size of hidden fully connected layers} & 100 to 2000 neurons\\ 
\emph{Activation for hidden layers} & ReLU~\cite{nair2010rectified}\\ 
\emph{Size of output layer} & 32 neurons\\ 
\emph{Activation for output} & Leaky ReLU~\cite{maas2013rectifier}\\ 
\emph{Optimizer} & Stochastic Gradient Descent~\cite{bottou2010large}\\ 
\emph{Dropout} & 0.1\\ 
\emph{Learning rate} & 0.001\\ 
\emph{Batch Size} & 512 pairs\\ 
\emph{Distance Metric} & Euclidean distance\\ 
\emph{Contrastive Loss Margin} & 10\\ 

\end{tabular}
}
\caption{Hyperparameters of our embedding neural network.}
\label{tbl:model}
\exhyphenpenalty 10000
\end{table}

For our classifier, we used the \textit{k-nearest neighbours} algorithm with $k=250$ for the first three experiments.
We were able to achieve better classification results by adjusting the $k$ parameter depending on the testing set
but $k=250$ produced consistently good results regardless of the number of classes. An advantage of maintaining the same 
configuration across all three of our experiments on webpage fingerprinting is that we can compare our findings more reliably.

\subsection{Exp. 1: Static Webpage Classification}
In this experiment, we assume an adversary that aims to fingerprint the pages of a small- or medium-sized website where all the pages share the same HTML template. This first experiment studies the performance of our proposed technique against a website with mostly-static webpages and a moderate percentage of shared content (the HTML template and the graphics).

\begin{figure}
    \centerline{
    \includegraphics[scale=0.85]{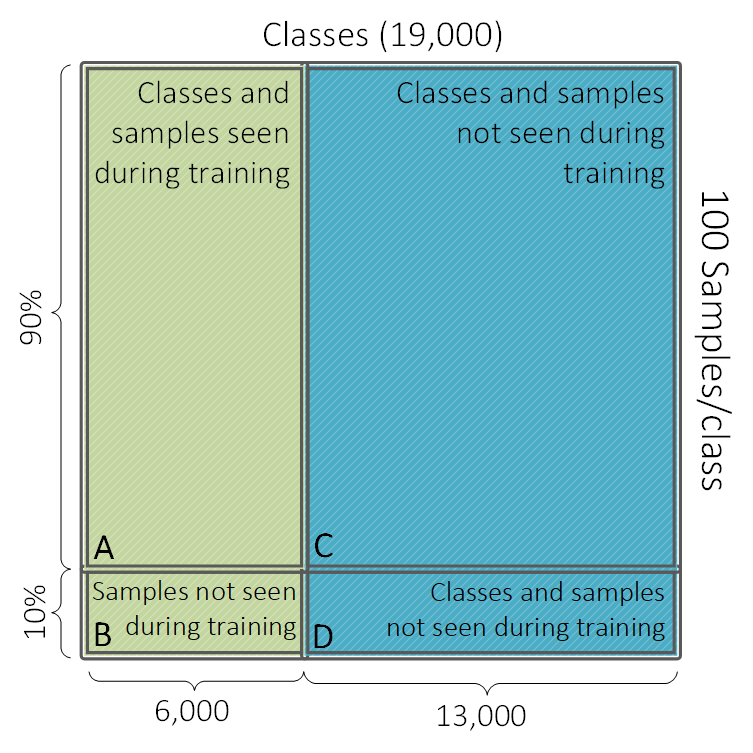}}
    \caption{For experiments 1 and 2, we use our Wikipedia dataset. The dataset is split into four smaller sets, both across its classes and its samples. Experiment 1 trains the embedding model on Set A and then validates the accuracy of the produced embeddings on previously-unseen samples from the same classes (Set B). In contrast, Experiment 2 reuses the trained model from Exp. 1 (trained on set A) to embed samples from Set C as reference points. Experiment 2 uses Set D as its \textit{test set}. Note that the classes in Sets C and D are not represented in sets A and B and vice versa. Moreover, no samples are shared between the Sets (e.g., no sequence from Set A is included in B, C or D).}
    \label{fig:wiki_set}

\end{figure}

Using our methodology from Section~\ref{sec:adaptive}, we train the adversary's embedding model on pairs of samples from our Wikipedia dataset. In particular, we use Set A (Figure~\ref{fig:wiki_set}) that includes 90 samples for each of the 6,000 distinct webpages/classes included in that Set. Upon completing the training phase, we deploy the model and use it to classify the samples in set B (Figure~\ref{fig:wiki_set}). The samples in set B originate from the same 6,000 
classes but correspond to traffic traces that were not used during the training phase (i.e., not included in Set A). During the classification phase, we use set A as the adversary's labeled sequences corpus (${\sim}$90 samples per class) and then use 
the trained model to classify the remaining ${\sim}$10 samples per class from set B (60,000 samples in total).

\begin{figure}
	\raggedleft
    \includegraphics[scale=0.48]{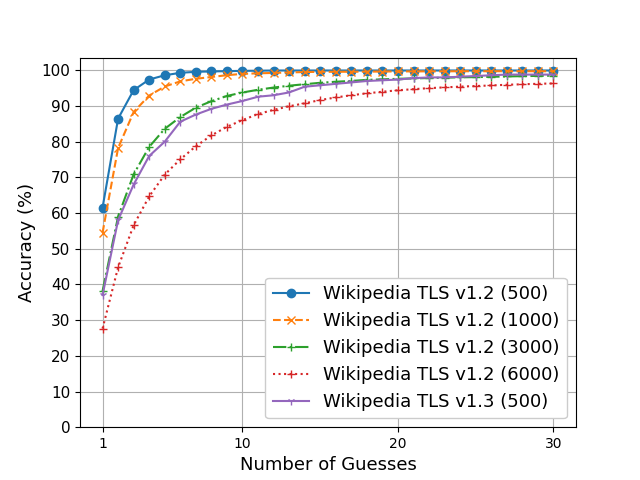}
    \caption{We evaluate the accuracy of the model in sets that required the adversary to attribute an encrypted traffic trace to a specific class from a set of 500, 1000, 3000 and 6000 possible Wikipedia articles (TLS 1.2). We also evaluate the same model on the 500-class set using TLS 1.3. For each class, we collected 100 samples, with 90 being used as reference points and the remaining 10 being classified by the model.}
    \label{fig:exp_comparison}
\end{figure}

To better study the performance of our model, we run our recognition task on different versions of Sets A and B containing 500, 1,000, 3,000 and 6,000 classes respectively. As seen in Figure~\ref{fig:exp_comparison}, out of a pool of 500 possible classes/articles, a top-3 adversary (i.e., the adversary is allowed to guess up to three classes) is able to correctly identify the Wikipedia article visited in more than 90\% of the cases. Moreover, top-1 adversaries have 58\% probability of correctly labeling the encrypted traffic trace, while top-10 adversaries are almost always able to correctly identify the page loaded.

In prior work~\cite{miller2014know}, Miller et al. used a hidden markov model (HMM) to take advantage of the adversary's prior knowledge of the website (i.e., which hyperlinks are available in each webpage) and fingerprint ``user journeys'' of up to 75 consecutive pageloads (through a set of 500 pages). They achieve an accuracy of 70-90\% depending on the length of the user journey fingerprinted. Our top-10 adversary has an accuracy of 55\% when fingerprinting from the set of classes with at least 25 samples each. The potential factors the impacted the performance are: the limited number of (reference) traces, the legacy TLS version (TLS 1.0) and the major change in the page structure. Exp. 3 investigates these factors further.

Moving on to larger sets, we evaluate the classification accuracy of our model in slices of Sets A and B with 1000, 3000 and 6000 classes (Figure~\ref{fig:wiki_set}). In the scenario of 1000 classes, a top-1 adversary is able to correctly classify previously unseen samples with 50\% accuracy, while in larger sets with 3000 and 6000 classes the same adversary achieves 35\% accuracy.
In the 1000- and the 3000-classes scenarios, the top-10 adversaries are able to correctly classify more than 90\% of the samples. In the 6000-classes case, a top-20 adversary also achieved above-90\% accuracy. In other words, an adversary who is allowed to choose 20 out of the 6000 labels ($0.3\%$ of the possible labels) has on average $>90\%$ likelihood of correctly inferring the page visited by the user. In this and the following experiment, the classification of a single example in the testing phase required $\leq$2 seconds. Adversaries that need to improve the performance further can easily parallelize the mapping and classification steps.

Overall, we demonstrated that adaptive fingerprinting adversaries are scalable and can classify with high accuracy samples originating from an extended pool of potential webpages. This result extends past works (\cite{danezis2009traffic,miller2014know}) on webpage fingerprinting that presented adversaries capable of classifying individual webpages and user journeys.

\subsection{Exp. 2: Adaptability \& Transferability}\label{exp:acrossclasses}
One of the goals of our methodology is to investigate whether an adversary can retain their classification accuracy even in cases of distributional shift (e.g., content changes, addition of new classes) at a minimal cost. Such a characteristic would significantly exacerbate the severity of fingerprinting attacks as it would 
make it practical to fingerprint a dynamic set of webpages where classes are added, changed and removed. 
Our fingerprinting methodology decouples these two tasks and allows the embedding model to remain 
class-agnostic, thus avoiding the need for any costly retraining. Instead, the adversary can easily
adapt to changes in the set of webpages or the contents of the webpages by updating the reference 
samples in the corpus of labeled traces.

To simulate a scenario of extreme distributional shift, we design an experiment where the adversary is classifying 
a set of articles that is completely disjoint from the set on which the model was trained, representing a worst-case 
scenario for an adversary. Such a difference between the training set and the testing set can occur 
in cases where the pages change drastically. For that purpose, we reuse the model trained in Experiment 1 
(on Set A) to embed samples in Sets C and D. As shown in Figure~\ref{fig:wiki_set}, Set A does not overlap with 
Sets C (and D) as the former contains samples from 6,000 classes while the latter contain samples from 13,000 \textit{different} classes.
We consider our testing set to comprise Sets C and D, where Set C populates the adversary's dataset of reference samples
and Set D contains the samples that need to be classified. As in Experiment 1, we investigate the accuracy of the model 
for slices of Sets C and D with different numbers of classes i.e., 500, 1000, 3000, 6000, 13000.

\begin{figure}
    \centerline{
    \includegraphics[scale=0.48]{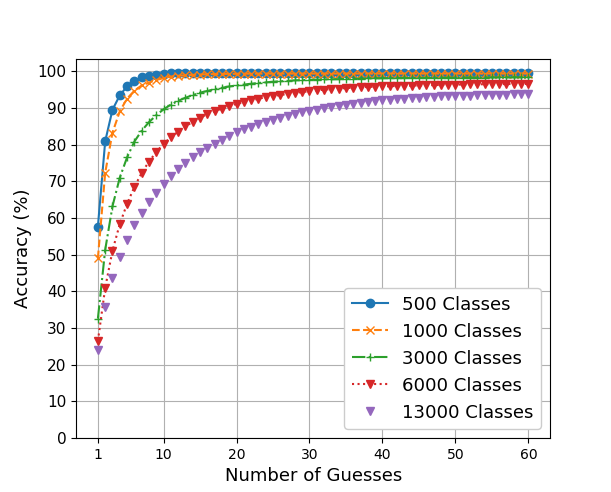}}
    \caption{Accuracy of our fingerprinting model for varying numbers of classes (Wikipedia articles) that were never encountered during training. The model was trained on a fixed set of 6000 Wikipedia articles and evaluated on a completely disjoint set of articles whose size ranged from 500 to 13,000 classes (both TLS 1.2). For each class, our dataset included 100 samples, with 90 being used as reference points and the remaining 10 being classified by the adversary.}
    \label{fig:exp_500-13000}
\end{figure}

As seen in Figure~\ref{fig:exp_500-13000}, the classification accuracy of the
adversary remains almost identical to the accuracy achieved with sets of the same 
size in Experiment 1 (i.e., without distributional shift). 
A top-1 adversary achieves 58\% accuracy in the 500-classes
set and a top-3 adversary ${\sim}$90\% accuracy. Similarly, a top-1 adversary 
achieves almost 50\% accuracy in the 1000-classes set and a top-4 adversary 
almost ${\sim}$90\% accuracy. 

This shows that the embedding model is learning the general leakage characteristics
of TLS streams rather than simply memorizing patterns that apply only to specific pairs 
of samples or classes from the training set. For example, through manual 
inspection of the traffic traces collected, we observed that the transmission patterns
of two samples from the same class can differ significantly. In one of them, the images 
were downloaded in multiple consecutive chunks of fixed length,
while in the other they were fetched as a whole. Despite these differences,
the model was correctly embedding the two samples in relative proximity.

Moreover, the adversary performs considerably well in even larger sets
of new classes. In particular, a top-10 adversary achieved an accuracy of 90\%,
80\% and 70\% in Sets with 3000, 6000, and 13000 classes respectively. 
This shows that our fingerprinting methodology can be reliably used to
embed and classify samples from classes that were never encountered during 
training.

As seen in Figure~\ref{fig:exp_500-13000}, 
the adversary needs to increase their number of 
guesses (i.e., parameter \textit{n} of a \textit{top-n} adversary)
as the number of classes increases in order for them 
to maintain the same level of accuracy (e.g., 90\%).
This is due to the increasing number of collisions between cross-class samples 
in the embeddings space. Intuitively, as the number of classes increases, 
the number of samples who are erroneously mapped in proximity to 
another class increases as well. However, as seen in 
Table~\ref{tbl:top-n}, \textit{n} increases slower than the 
number of classes. This implies that while the absolute number
of collisions increases with the number of classes, the increase 
in collisions has a sublinear relationship with the increase of the 
number of classes. In other words, for any percent increase
in the number of classes the adversary needs to increase
their $n$ by less than 1\%.

\begin{table}    \centering
    \caption{As the number of classes increases the accuracy of the embeddings decreases as cross-class collisions become more likely. However, \textit{n} has a sublinear relationship with the number of classes.}
    \begin{tabular}{cccc} 
        \hline
             \# Classes & Top-$n$ & Accuracy & $\frac{n}{\# \textrm{Classes}}$\% \\ \hline
            500 & 3  & 89\% & 0.6\%  \\
            1000 & 4  & 89\% & 0.4\%  \\
            3000 & 10  & 90\% & 0.33\%  \\
            6000 & 20  & 92\% & 0.33\%  \\
            13000 & 30  & 89\% & 0.23\%  \\ \hline \\
        \end{tabular}
    \label{tbl:top-n}
\end{table}

Collisions might also occur if samples from classes not in the ``labelled traces'' were captured by the adversary. During classification such an ``unknown'' pageload may be an obvious outlier (i.e., no proximity to any of the known labels in embeddings space) or may  with an known class (leading to a misclassification). 
To avoid this, the adversary has to keep track of the website's pages (e.g., through sitemaps, crawling) and include any new pages in their reference set.

\subsection{Exp. 3: TLS Version \& Theme Sensitivity}\label{sec:version}
In this experiment, we examine the learning characteristics of our
adaptive fingerprinting adversary. In particular, we evaluate the 
effect of retaining multiple IP sequences, and the degree distributional shift affects fingerprinting across websites and TLS versions.
To determine the sensitivity of our trained fingerprinting model to the TLS version, we evaluated the model from Exp.1 on a set of 500 TLS 1.3 Wikipedia webpages. These webpages were seen during training but only through TLS 1.2 traffic traces. As seen in Figure~\ref{fig:exp_comparison}, switching to TLS 1.3 has an impact on the model's accuracy with the top-3 adversary achieving an accuracy of approximately 70\% while on TLS 1.2 the top-3 adversary achieves approximately 95\%. The adversary still retains some of its accuracy hinting that the model learned some protocol version-specific features but does not rely entirely on them.

To determine the impact of switching to a completely different website (both by TLS version and theme), we fingerprint traces from Github README dataset (TLS 1.3, 500 webpages from the top 500 open-source projects, Section~\ref{sec:datasets}). However, Wikipedia pageloads always involve 3 IP addressed (i.e., the client's browser, text server media server), while Github pages load resources from a non-constant number of servers. As our model operates on a fixed number of sequences, we opted to represent the traffic as two sequences (i.e., traffic from and towards the user's browser). For this reason, we could not reuse our model from Experiment 1 (as it is trained to process three sequences) and had to retrain it to work on two sequences. We then ran the recognition task again on the original Wikipedia dataset (for a baseline) and on the Github dataset (both represented as two sequences).

\begin{figure}
    \centerline{\includegraphics[scale=0.48]{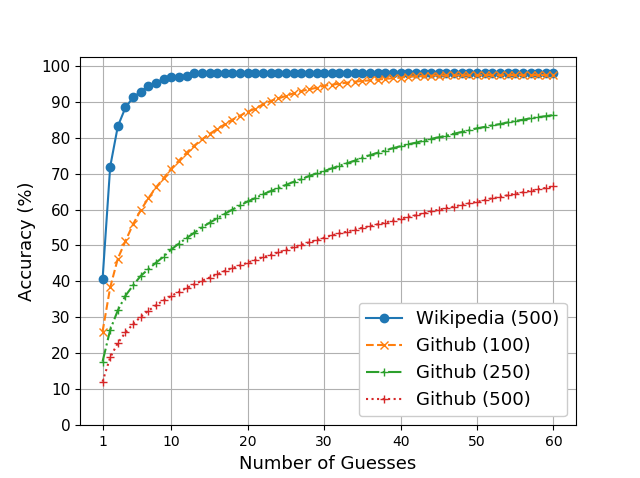}}
    \caption{We trained our embedding model on two-sequence traffic traces from Wikipedia (TLS 1.2) and used it to embed and classify traces collected from Github (TLS 1.3). The model performs considerably better when operating on traces from the same website and with the same protocol version it was trained on, however, it still retains some of its accuracy. This indicates that some leakage characteristics are preserved even across very different setups.}
    \label{fig:exp_github}
\end{figure}

The performance of the model on the three versions of the Github dataset (i.e., Github 100, 250, 500) shows that adversaries are able to retain a fair classification accuracy even in this case of extreme distributional shift across multiple dimensions. This indicates that some leakage characteristics persist across IP encoding, websites and protocol versions. Nonetheless, the reduced accuracy 
hints that the embedding model is sensitive to distributional shift with changes in the website's theme having the most impact.

\subsection{Exp. 4: Class Distinguishability}
The previous experiments report the accuracy of our adversary over all the fingerprinted classes. However, an attacker might attribute higher value to only some sensitive pages that may reveal the political or religious affiliations of the user. As discussed in~\cite{RimmerPJGJ18}, there can be no definitive set of sensitive webpages/websites, however, we can study how accurate the adversary is on a per class basis. This is of particular interest as two adversaries with identical accuracy over a set of pages A, might have very different performance when we consider only the subset of sensitive pages out of A. Note that the proportions shown here are calculated on a per class basis, unlike the accuracy curves from Section~\ref{sec:experiments} which are calculated on a per sample basis. The difference is that the per-class mean is sensitive to outlier samples.

\begin{figure}
    \centerline{
    \includegraphics[scale=0.48]{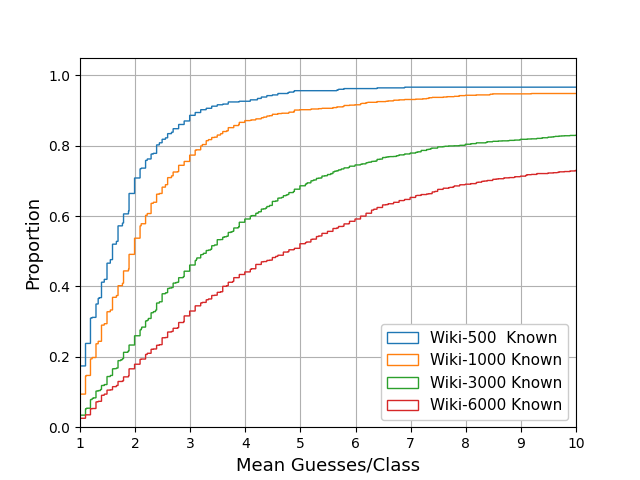}}
    \caption{Cumulative distribution of the mean number of guesses needed (per class) for traces from webpages seen during training.}
    \label{fig:cumul-known}
\end{figure}

\begin{figure}
    \centerline{
    \includegraphics[scale=0.48]{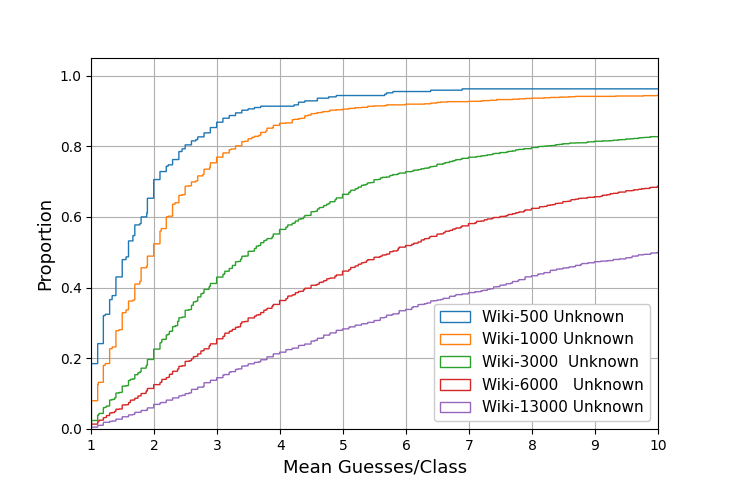}}
    \caption{Cumulative distribution of the mean number of guesses needed (per class) for traces from webpages not seen during training.}
    \label{fig:cumul-unknown}

\end{figure}

\begin{figure}
    \centerline{
    \includegraphics[scale=0.48]{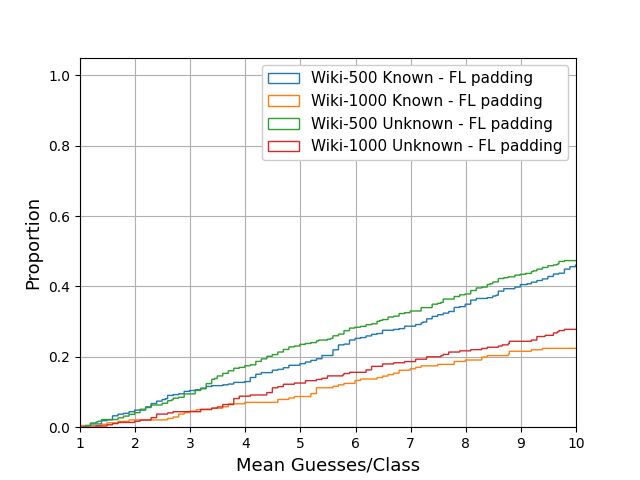}}
    \caption{Cumulative distribution of the mean number of guesses (per class) for padded traces. We test both on datasets with classes that were used during training and classes that were not used.}
    \label{fig:cumul-padding}
\end{figure}

We use the trained model from Experiment 1 (Section~\ref{sec:experiments}) on our Wikipedia dataset. Figures~\ref{fig:cumul-known},\ref{fig:cumul-unknown},\ref{fig:cumul-padding} show the cumulative distribution of the guesses that the adversary had to make (top-N accuracy) per class. We report the average number of guesses per class (cf. per sample) as we want to study if certain pages are easier to fingerprint than others. 

As in Experiments 1 and 2, we don't observe major differences between the model tested on known classes and unknown ones. In both cases, there is a substantial percentage of classes (e.g., 40\% for $<$2 guesses in Wiki-500 and Wiki-1000) that the model does exceptionally well against. In contrast, there is a small percentage of classes ($\sim$3\%) that our adversary finds hard to distinguish even in small sets (Wiki-500, Wiki-1000).
However, the differences in the distinguishability of the classes should not be attributed solely on the contents of the pages or on models' bias. In practice, these two factors are probably intertwined with distinctive pages being more distinguishable and the model training process favouring updates that result in immediate loss reduction (over harder to learn features). This becomes relevant when considering defences, as easier-to-learn features are not necessarily the most robust to countermeasures.

We now evaluate whether padding countermeasures can adequately protect even those more distinguishable classes. 
Using the fixed length (FL) padding setup, we plot the cumulative distribution over the average guesses per class. Figure~\ref{fig:cumul-padding} shows that fixed length padding substantially reduces the number of classes that are distinguished within a few guesses (i.e., low N in top-N adversary). In fact, the proportion of the classes that have been correctly inferred after 10 guesses (N=10) is smaller than the proportion of classes for N=1 in a non-padding scenario. It follows that FL padding is indeed effective against adaptive adversaries and more importantly is effective even on pages that are normally easy to distinguish.

\section{Discussion on Countermeasures}\label{sec:defenses}

Adaptive fingerprinting 
attacks can affect both the users of anonymity networks
and the users of the TLS protocol (i.e., a very large portion 
of the Internet users). However, the scope of potential defenses 
for the TLS protocol is limited to only those countermeasures 
that have only a very light impact on the bandwidth used.
Intuitively, a protocol-level countermeasure with a 10\% 
bandwidth overhead, would result in an approximately equal 
increase in the web-traffic bandwidth worldwide.
For this reason, the majority of the defenses proposed for Tor are not directly applicable to TLS.

As specified in Section~\ref{sec:threatmodel}, \textit{webpage} 
fingerprinting aims to infer the specific page visited by a user
from a set of pages all of which belong to the same website. This
is a major difference to the website fingerprinting setup.
In particular, each website can be treated as a separate entity
and thus the defenses can be deployed and adjusted on a per-website basis.
For example, a website with non privacy-sensitive pages (e.g., a list of
hardware drivers) could decide to not deploy any countermeasure or optimize the
deployment for low bandwidth impact (cf. for privacy).
On the other hand, a website with sensitive content could 
use a more conservative configuration. 

Being able to configure the countermeasure on a per-website basis, allows us to achieve protection without increasing the bandwidth overhead disproportionately. In comparison, defenses for website fingerprinting attacks rely on a cross-website anonymity set and thus require the deployment of the specific countermeasure by several websites in order to be effective.

For example, such a per-website policy could use padding to conceal the byte length of the webpages loaded. This approach conceals not only the length of each individual transmitted packet but also prevents timing attacks (e.g., with additional dummy packets). An advantage of this approach is that TLS already has this capability and thus would not require any protocol changes~\cite{rescorla2018transport, pironti2013length}. Moreover, given that padding is a well-studied technique, we could draw useful lessons from prior works in the area (e.g., Pironti et al.~\cite{pironti2012identifying} have shown that random-length padding is not sufficiently effective).

\begin{figure}
    \includegraphics[scale=0.48]{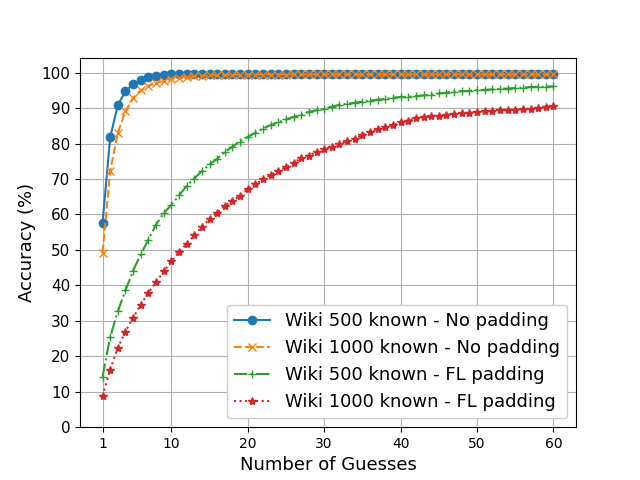}
    \caption{We test our adversary on 500 and 1000 classes that were seen during training (known). Despite this being a favourable scenario for the adversary, we observe that fixed length (FL) padding significantly obfuscates the origin of a given trace making it hard for the embedding model to retain its accuracy.}
    \label{fig:padding_known}
\end{figure}

\begin{figure}
	\includegraphics[scale=0.48]{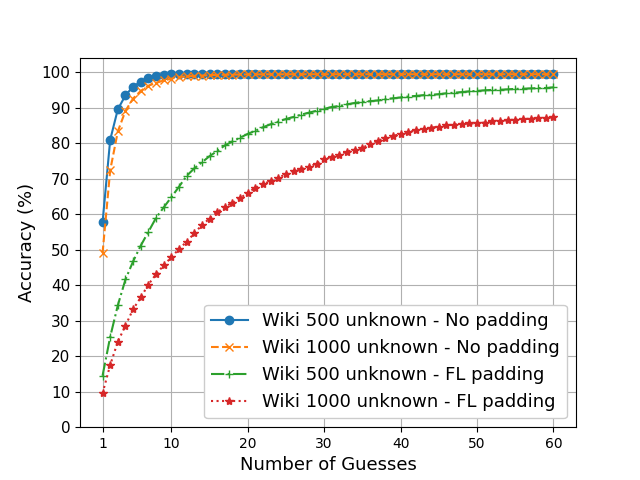}
	\caption{We test our adversary on a less favourable scenario with 500 and 1000 classes that were not used during training (unknown). We see that fixed length (FL) padding significantly affects the accuracy of the model.}
	\label{fig:padding_unknown}
\end{figure}

Based on our Wiki19000 dataset, we now evaluate the effectiveness of fixed length padding against
the model used in experiments \#1 and \#2. We apply fixed length (FL) padding as suggested in the TLS specification~\cite{rescorla2018transport}. More specifically, given a set of target webpages, 
we padded all the traces to match the length of the longest one. Figure~\ref{fig:padding_known} 
compares the performance of the adversary on the non-padded traces with their performance on the 
padded traffic. We used traces from 500 and 1000 classes that were used when training the model.
We see that there is a noticeable decrease in the performance of the adversary but not a complete loss of accuracy. This hints that the embedding model was still able to extract some useful features despite the additional noise. Similarly, Figure~\ref{fig:padding_unknown} shows the performance of the adversary on traces from 500 and 1000 classes that were \textit{not} used during the training. We observe that again padding was able to significantly decrease the performance of the adversary.

In the above experiment, we were interested in measuring our attack against one of the strongest possible defenses. We observe that, indeed, the FL padding mechanism significantly impacts the accuracy of our TLS fingerprinting adversary. However, as noted in~\cite{6234422} general-purpose countermeasures are unlikely to be bandwidth-efficient, which is true also for FL padding.

To alleviate this overhead, we could relax the requirement for complete indistinguishability between all the webpages of a website, and specify a padding policy that provides adequate privacy without impractical bandwidth overheads. For example, each website operator could configure their selected countermeasure so as to always guarantee a minimum anonymity set size (i.e., number of webpages that are indistinguishable) depending on how privacy-sensitive the pages of the website are. We expect that smaller websites ($<500$ webpages) could make all their pages indistinguishable at a relatively low bandwidth cost, while websites with more pages (e.g., Wikipedia) will have to split their content into smaller anonymity sets and aim for intraset indistinguishability. Such a strategy could be combined with FL padding or other padding techniques (e.g., Adaptive Padding~\cite{juarez2016toward}). Another potential direction would be to combine privacy-preserving techniques (e.g.,~\cite{abs220303764,hicks2018vams}) to mix traffic from multiple pages but this would impose a significant bandwidth usage increase.

\section{Related Work}\label{sec:relatedwork}
\begin{table*}
	\centering
	\caption{Comparison of the operational costs between various different fingerprinting approaches. Some fingerprinting models were evaluated on a range of values. We use ``?'' when information was not reported by the authors.}
	\label{tbl:costs}
	\begin{tabular}[t]{lccc|cc|cc}
		\multicolumn{4}{c}{\textbf{System}} & \multicolumn{2}{c}{\textbf{Training}} & \multicolumn{2}{c}{\textbf{Update}}\\
		\cline{1-8}
		\textit{Name} & Protocol & Classes & D. Shift & Instances & Complexity & Retraining & Instances
		\tabularnewline
		\tabularnewline
		\textbf{Adaptive Fingerprinting} & TLS & up to 13.000 & \cmark & 90 & High & \xmark & 90\\
		Miller et al.~\cite{miller2014know} & TLS & 500  & \xmark & 1-200 & Moderate  & \cmark & 1-200\\
		Bissias et al.~\cite{bissias2005privacy} & SSL & 100 & \xmark & ? & Low & \xmark & ? \\
		Triplet Fingerprinting~\cite{sirinam2019triplet} & Tor & up to 775 & \cmark & 25 & High & \xmark & 5-20 \\
		Deep Fingerprinting~\cite{sirinam2018deep} & Tor & 95 & \xmark & 1000 & High & \cmark & 1000\\
		Var-CNN~\cite{bhat2019var} & Tor & up to 900  & \xmark & 10-1000 & High & \cmark & 10-1000\\
		k-fingerprinting~\cite{hayes2016k} & Tor & up to 100 & \xmark & 60 & Moderate & \xmark & 60\\
	\end{tabular}
\end{table*}%

Various insights used in webpage fingerprinting papers were motivated by works on \textit{website} fingerprinting attacks against the Tor anonymity network~\cite{dingledine2004tor}. Such attacks infer the website that the user has visited but not the specific webpage loaded~\cite{wang2014effective, panchenko2016website, sirinam2018deep, hayes2016k, bhat2019var}. Previous works on Tor-based website fingerprinting have employed standard machine learning techniques for classification such as k-NN~\cite{wang2014effective}, Support Vector Machines~\cite{panchenko2016website}, random forests~\cite{hayes2016k}, and more recently neural networks~\cite{sirinam2018deep, bhat2019var}.
To better understand the operational cost our adaptive fingerprinting adversary incurs, we compare it with previous works using the framework introduced in~\cite{juarez2014critical}. We note that webpage and website fingerprinting are very different tasks and trained fingerprinting models cannot be transferred between the two (e.g., our model relies on knowledge of the IP addresses involved in a page load). However, comparing the computational and operational costs of our approach with state-of-the-art webpage and website fingerprinting designs is possible, as the underlying techniques might be transferable to some extent.

Table~\ref{tbl:costs} presents the operational costs of our design and some of the most notable designs from the webpage and website fingerprinting literature. The cost modeling framework introduced in~\cite{juarez2014critical} defines a \textit{training}, \textit{testing} and \textit{updating} tasks as well as a \textit{data collection} task. In practice, data collection is used as a sub-task to fetch the data needed for the other three tasks. 
According to the framework, the collection cost of a dataset D (\textit{col}(D)) depends on the cost of fetching one sample (\textit{col}(1)) and the number of pages the adversary needs to collect $D=n \times m \times i$, where $n$ is the number of webpages/websites/classes, $m$ the average number of versions of each webpage/website that are different enough to reduce the classifier’s accuracy, and $i$ is the number of instances per webpage/website/class. Parameters $n$, $m$ and \textit{col}(1) depend on the website(s) fingerprinted and the crawler used, and not on the fingerprinting model. However, $i$ depends on the amount of data required by the fingerprinting model. For a fair comparison, we assume that all the models are evaluated on the same dataset collected with the same crawling software, and thus focus on parameter $i$ which is model-dependent.

The training cost is defined as $col(D) + train(D,F,C)$
, with $D$ being the data needed for training, $F$ the cost of measuring/extracting features and $C$ the cost of training the classifier. As seen in Table~\ref{tbl:costs}, systems based on low and moderate complexity models (k-fingerprinting~\cite{hayes2016k}, Miller et al.~\cite{miller2014know}) require fewer instances ($i$) per page/class, while models based on deep neural networks (e.g., Deep Fingerprinting) typically require significantly more instances/data. However, more recent works (e.g., Var-CNN~\cite{bhat2019var}, Triplet fingerprinting~\cite{sirinam2019triplet}, ours) took advantage of the advances in ML training algorithms to again reduce the number of instances needed, even when complex models are  trained. Nevertheless, all neural network-based approaches (e.g., ours, Triplet Fingerprinting~\cite{sirinam2019triplet}, Var-CNN\cite{bhat2019var}), require a computer with a capable Graphics Processing Unit card and enough RAM, thus entailing an increased training cost ($train(D,F,C)$). Given that each model is trained only once (provisioning phase in our case), we consider this cost to be manageable even by less well-resourced adversaries.

Similarly, the testing cost is defined as $col(T) + test(T,F,C)$, where $T=v \times p$ with $v$ being the number of victim users and $p$ the average number of pages loaded by each victim per day. As above, we fix $v$ and $p$ to be the same for all the models compared. Thus, the testing cost depends entirely on the cost of extracting the features given a single trace and the cost of using the classifier. Using a trained model is not a computationally expensive task even for complex models (e.g., a neural network). Most works report a time of 1-2 seconds per sample inference. We also measured times of $\leq$2 seconds per sample.

Finally, to maintain the performance of the classifier, the adversary needs to update the system over time. The frequency of those updates depends on the nature of the website(s) fingerprinted. In this case, we observe significant differences between the systems studied.
The total cost of an update is: $col(D) + update(D,F,C)$, which includes the cost of updating the data (D), measuring the features (F) and retraining the classifier. In systems that use simple models (e.g., k-fingerprinting~\cite{hayes2016k} and Bissias et al.~\cite{bissias2005privacy}), the only cost is that of collecting new samples and extracting their features (no need for retraining after the initial calibration). However, the accuracy of such models in moderate and large sets has not been shown. More capable systems such as Deep Fingerprinting~\cite{sirinam2018deep} and Var-CNN~\cite{bhat2019var} require to be retrained with each update and thus incur substantial computational costs on top of the data collection costs. 

Models based on embeddings do not require to be retrained when updating. The adversary can swiftly swap the samples in the reference dataset with new ones so as to keep up with content updates or to include additional webpages/websites in the set. This process does not involve any retraining as the embedding model can process any traffic trace even if it originates from a class not encountered during training. This simplifies the update process to only a few low-complexity operations (i.e., collecting and embedding new samples) and enables the adversary to easily keep up with any distributional shift. 
Overall, classifiers based on embedding models combine both the low operational costs of simple systems with the scalability of neural networks. In comparison, all past works on webpage fingerprinting assume a non-changing target set and would require some form of retraining to keep up with changes in the input distribution~\cite{danezis2009traffic,bissias2005privacy,cai2012touching,miller2014know}. While this cost may seem reasonable when considering small, fixed target sets ($<500$ webpages), it quickly grows (due to the constant retraining required) when considering hundreds or thousands of changing pages.

Note that this work does not consider active attacks where adversaries can drop, inject or delay packets. Such techniques could result in even greater privacy loss due to the
adversary's ability to force targetted re-transmissions of parts of the webpage loaded.
We consider reinforcement learning~\cite{sutton2018reinforcement} a suitable paradigm for modeling such adversaries. For example, the PPO algorithm~\cite{schulman2017proximal} has already found numerous applications in cybersecurity~\cite{cage3aicd,foley,bottinger2018deep,cage1_win,cage2aicd} and could be adapted to explore for efficient ``active'' fingerprinting strategies.

\section{Conclusions}
This work focuses on modern webpage-fingerprinting adversaries and studies their performance under realistic conditions and constraints. We show that they are indeed effective against TLS even under non-optimal conditions and discuss how appropriate defenses can be deployed. An interesting direction for future work would be to evaluate whether VPN services are vulnerable to adaptive adversaries (the task is to predict the domain name, not the full URL). Another future research direction is to measure the empirical privacy offered by CDNs in the wild, and whether it can be circumvented by adversaries (e.g., CDN latency as side-channel). 

\section{Acknowledgements}
We extend our gratitude to our colleagues Shehar Bano and Kostas Papagiannopoulos for generously providing their insights and feedback during the initial phases of this project.
\newpage
\bibliographystyle{IEEEtran}
\bibliography{biblio}

\end{document}